\documentclass[pre,aps,amsmath,amssymb,twocolumn,showpacs,floatfix,superscriptaddress,
longbibliography]{revtex4-1}
\usepackage{graphicx}
\usepackage{dcolumn}
\usepackage{bm}
\usepackage{mathtools}

\begin{document}

\title{Resonant Anderson localization in segmented wires} 

\author{Cristian Estarellas}
\affiliation{Institut de F\'{\i}sica Interdisciplin\`aria i de Sistemes Complexos IFISC (CSIC-UIB), E-07122 Palma de Mallorca, Spain}
\author{Lloren\c{c} Serra}
\email{llorens.serra@uib.es}
\affiliation{Institut de F\'{\i}sica Interdisciplin\`aria i de Sistemes Complexos IFISC (CSIC-UIB), E-07122 Palma de Mallorca, Spain}
\affiliation{Departament de F\'{\i}sica, Universitat de les Illes Balears, E-07122 Palma de Mallorca, Spain}

\date{October 17, 2015}

\begin{abstract}
We discuss a 
model of random segmented wire, with linear segments of 2D wires joined
by circular bends. The joining vertices act as scatterers
on the propagating electron waves. The model leads to 
resonant Anderson 
localization when all segments are of
similar length. The resonant behavior
is present with one and also with several propagating modes.
The probability distributions evolve from diffusive to localized regimes 
when increasing the number of segments
in a similar way for long and short localization lengths.
As a function of the energy a finite segmented wire typically evolves from localized to diffusive to ballistic behavior in each conductance plateau.
\end{abstract}


\maketitle

\section{Introduction}

The electrical transport properties of bend semiconductor  nanowires attracted much interest some years ago \cite{Wu,Sprung,Sols,Lent}. In particular, phenomena such as the formation of localized states and the 
scattering behavior of circular bends were considered.
At low energies it was proved that a circular bend can be understood as an attractive square well
supporting bound states. 
Recently, we extended a similar analysis to closed polygons made of 2D nanowires,
finding characteristic sequences of eigenstates \cite{Estarellas15}.
The optical absorption of polygonal nanorings has also been recently  considered in Ref.\ \cite{Sitek15}.

\begin{figure}[t]
\centering
\resizebox{0.4\textwidth}{!}{
	\includegraphics{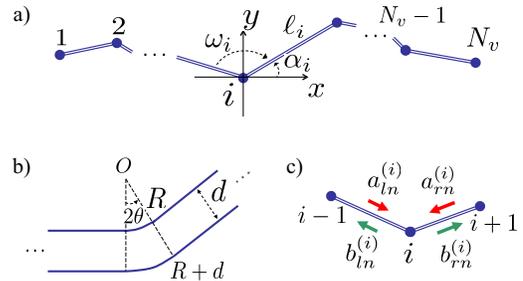}
}
\caption{a) Sketch of a segmented 2D planar nanowire showing the 
definitions of the random angles $\alpha_i$ and $\omega_i$ for each vertex and the 
segment lengths $\ell_i$.
b) Model of the vertex as a circular bend, with the definitions 
of the bend radius $R$ and angle $2\theta$.
c) Notation for the scattering amplitudes for left and right incidence on vertex $i$.
}
\label{F1}
\end{figure}

Motivated by the above studies we present in this work a model of a segmented wire,  
made
with a large collection of straight segments joined with circular bends (Fig.\ \ref{F1}). We address the localization properties in the resulting nanowire when the vertices (bends) and segment lengths vary randomly. It is well known that the quantum interference of scattered waves in presence of disorder leads to the phenomenon of Anderson localization \cite{Anderson58}.
Reviews on this long-lasting topic with an extensive literature are,
e.g.,
Refs.\ \cite{Beenakker97,Kramer93,Pendry94,Deych01,Datta,Mello}.
We address in this work the localization phenomenology of the segmented wire model.

Anderson localization in disordered 1D systems has been extensively investigated. In particular, the model
of successive 1D barriers (or wells) of fixed thickness $\ell$ and random heights is known to lead to resonant localization whenever the accumulated phase in the distance $\ell$ is an integer multiple of $\pi$ \cite{Diaz12,Herrera13,Diaz15}. We show below that the segmented wire exhibits resonant localization when
the segment lengths are narrowly distributed around a given value $\ell_0$. This resonant  localization occurs not only in the regime of one propagating mode, where one normally expects the fully 1D 
behavior, but also in regions with several propagating modes.
 
Localization in quasi-1D systems has been considered in wires with bulk and surface disorder 
(see, e.g., Ref.\ \cite{Froufe10} and references therein).  
Indeed, analytical models based on Fokker-Planck equations have been developed for both types of disordered waveguides \cite{Dorokhov82,Mello88,Froufe07}, as well as field-theoretic
equivalent approaches \cite{Efetov83,Brouwer96}. 
A disorder-to-chaos transition when varying the degree of edge corrugation of a quasi-1D waveguide has been recently predicted in Ref.\ \cite{Alcazar13}. 
Quasi-1D tight binding and analytical models with correlated disorder have been discussed in Refs.\ \cite{Herrera14,Izrailev05} and cylindrical shells were studied in Ref.\ \cite{Serra09}. Those studies, however, did not focus on resonant behavior of the corresponding quasi-1D models.
Our work is an approach to the resonant localization of the segmented wire, emphasizing  
the physical modeling of the 2D wire bends, complementing more
methodological approaches focusing on the fundamentals of localization 
theory. 

The paper is organized as follows. Section II presents the model, briefly describes the transport problem and the calculation of the vertex scattering matrix. Section III contains the definitions of the localization properties and Sect.\ IV contains the results. Finally, the conclusions of the work are summarized in Sect. V.

\section{Segmented wire model}

\subsection{Model definition}
We consider a quantum waveguide formed by joining straight 2D channels
of width $d$. The joining vertices are described by simple circular bends
of radius $R$ and angle $2\theta$. Figure \ref{F1} sketches the system
definitions. A set of values $\{\ell_i,\alpha_i,R_i,i=1,\dots,N_v\}$, with 
$N_v$ the number of vertices,
gives a particular physical realization of the segmented wire. We are interested in the statistical 
properties when these values vary randomly within given intervals
\begin{eqnarray}
\alpha_i &\in& [-\alpha_0,\alpha_0]\; ,\\
\ell_i &\in& [\ell_0(1-p_\ell),\ell_0(1+p_\ell)]\; ,\\
R_i &\in& [R_0(1-p_R),R_0(1+p_R)] \; ,
\end{eqnarray}
where $\alpha_0$, $\ell_0$, $R_0$, $p_\ell$ and $p_R$ 
characterize the parameter ranges of random variation. The parameters 
$p_{\ell}$ and $p_{R}$ are dimensionless and fulfill $0\le p_{\ell,R}\le 1$. They 
represent the maximum random variation (in relative terms) 
of the segment lengths around $\ell_0$ and of the vertex radius 
around $R_0$, respectively.

\subsection{The transport problem}
Each vertex is characterized by a scattering matrix relating input and output wave amplitudes
\begin{equation}
\label{eq4}
\left(
\begin{array}{c}
b_{l}^{(i)}\\
b_r^{(i)}
\end{array}
\right) 
=
\left(
\begin{array}{cc}
r^{(i)}& t^{(i)}\\
t^{(i)}& r^{(i)}
\end{array}
\right) 
\left(
\begin{array}{c}
a_{l}^{(i)}\\
a_r^{(i)}
\end{array}
\right)\; ,
\end{equation}
where the notation of Fig.\ \ref{F1}c is used. For the case of multiple propagating modes,
input and output amplitudes in Eq.\ (\ref{eq4}) correspond to vectors.
For instance,
$a^{(i)}_ l\equiv (a^{(i)}_{l1},a^{(i)}_{l2},\dots,a^{(i)}_{lN_p})$, 
with $N_p$ the total number of 
propagating modes.
Analogously, the transmission and reflection coefficients become matrices,
$t^{(i)}\equiv t^{(i)}_{nn'}$,
$r^{(i)}\equiv r^{(i)}_{nn'}$ with $n,n'=1,\dots,N_p$.

There is a relation between input and output amplitudes for successive vertices. For instance,
right-output from vertex $i-1$ in mode $n$ coincides with left-input for vertex $i$, with a phase, i.e., 
$b^{(i-1)}_{rn}=a^{(i)}_{ln}\exp{(-i k_n l_{i-1})}$. A closed system of linear equations 
is obtained assuming unit left incidence on vertex 1 in mode $n_{i}$. The linear system of $4N_pN_v$ equations reads

\begin{widetext}
\begin{equation}
\begin{array}{rcll}
\rule{0cm}{0cm} a^{(i)}_{ln} &=& \delta_{n,n_{i}} \quad & (i= 1)\; , \\
\rule{0cm}{0.55cm} a^{(i)}_{rn} &=& 0 \quad & (i=N_v)\; ,\\
\rule{0cm}{0.55cm} b^{(i)}_{ln}-a^{(i-1)}_{rn}\exp{(-i k_n l_{i-1})} 
 &=& 0\quad & (i\ne 1) \; ,\\
\rule{0cm}{0.55cm} b^{(i)}_{rn}-a^{(i+1)}_{ln}\exp{(-i k_n l_i)} 
 &=& 0\quad & (i\ne N_v) \; ,\\
\rule{0cm}{0.55cm} b^{(i)}_{ln}
-\sum_{n'}{r^{(i)}_{nn'} a^{(i)}_{ln'}}
-\sum_{n'}{t^{(i)}_{nn'} a^{(i)}_{rn'}}
 &=& 0 \quad & (i= 1,\dots,N_v)\; ,\\
\rule{0cm}{0.55cm} b^{(i)}_{rn}
-\sum_{n'}{t^{(i)}_{nn'} a^{(i)}_{ln'}}
-\sum_{n'}{r^{(i)}_{nn'} a^{(i)}_{rn'}}
 &=& 0 \quad & (i= 1,\dots,N_v)\; ,
\end{array}
\end{equation}
\end{widetext}
where $n$ and $n'$ are mode indexes ranging from 1 to $N_p$.
The solution of the linear system yields the output amplitudes on each vertex as a function of
the incidence mode $n_i$, $b^{(i)}_{rn}(n_i)$.
The total transmission is obtained by adding the modulus squared of the right-output amplitudes
on the last vertex $N_v$  as  
\begin{equation}
T=\sum_{n_{i}n_{o}}{\left| b^{(N_v)}_{rn_o}(n_i) \right|^2}\; .
\end{equation}

Once the scattering matrices are known, the linear system can be efficiently solved in a numerical way for quite large values of the number of segments $N_v$. The  computer solution is also fast enough to allow for an statistical analysis with the 
random variation of the parameters (Fig.\ \ref{F1}a). 

\subsection{Vertex scattering matrix}

The transmission and reflection matrices, $t$ and $r$, for a single vertex are a required input of the model. We describe each vertex as a circular bend of the 2D wire (Fig.\ \ref{F1}b), a problem that was 
studied in Refs.\ \cite{Wu,Sprung,Sols,Lent}. In particular, we have followed the approach of Ref.\ 
\cite{Sols} that relies on the separability of Schr\"odinger's equation in the bend region in radial and angular parts. Using polar coordinates for the bend region of Fig.\ \ref{F1}b, with $\rho$ the 
distance to $O$ and $\phi$ the azimuthal angle, the wave function may be factorized as 
$P_{|\nu|}(\rho)\exp{(\pm|\nu|\phi)}$. 
The eigenvalue problem given in Eq.\ (2b) of Ref.\ \cite{Sols}
determines both  $\nu^2$ and 
$P_{|\nu|}(\rho)$. For completeness, we repeat here this eigenvalue problem
\begin{eqnarray}
\rho^2 \frac{d^2 P_{|\nu|}(\rho)}{d\rho^2}
+\rho\frac{d P_{|\nu|}(\rho)}{d\rho}
&+& \frac{2mE}{\hbar^2}\rho^2 P_{|\nu|}(\rho)
=\nonumber\\
&-&\nu^2 P_{|\nu|}(\rho)\; .
\label{eq11}
\end{eqnarray}

Equation (\ref{eq11}) is not apparently Hermitian. However, the transformation 
$P_{|\nu|}=\sqrt{\rho}F_{|\nu|}$ leads to
\begin{equation}
\label{eq12}
\rho \frac{d^2}{d\rho^2}\rho\, F_{|\nu|}(\rho)
+\left(\frac{1}{4}+\frac{2mE}{\hbar^2}\rho^2\right) F_{|\nu|}(\rho)
=
-\nu^2 F_{|\nu|}(\rho)\; ,
\end{equation}
that, being Hermitian, can be numerically diagonalized with standard routines for symmetric matrices.
As $\nu^2$ is thus a real value, $\nu$ can be either real or purely imaginary, which describe evanescent and propagating  angular waves in the bend, respectively. Determining this way the bend modes, the matching conditions at the interfaces (dashed lines of Fig.\ \ref{F1}b) 
yield the required scattering matrices. 
As mentioned in  Ref.\ \cite{Sols}, 
truncating the number of bend modes and straight wire modes to the same value, a linear 
system of equations replaces the matching conditions. We have implemented that method and checked that our solution reproduces 
the transmission and reflection probabilities given in Ref.\ \cite{Sols}.

At low energies it was shown in Ref.\ \cite{Sprung} that the circular bend may be approximated by a 1D square well of depth $V_0=\hbar^2/(2m\bar{R}^2)$  and width $2a=2\bar{R}\theta$, where 
$\bar{R}=\sqrt{R(R+d)}$ is an average effective radius. In this approximation the scattering matrices are, of course, analytical. 
 
\section{Localization properties}

The localization length $\ell_{\it loc}$ is a characteristic distance
such that random segmented wires whose total length $L$ fulfills $L>>\ell_{\it loc}$ present a
statistical distribution of $\log(T)$ that is Gaussian (normal) distributed around a mean value.
Of course, the localization length depends on the system parameters as well as on the energy $E$
of the transport electrons.
Deeply in the localized regime ($L>>\ell_{\it loc}$) the transmission is in general greatly quenched for 
the huge majority of system realizations. It is therefore very relevant to characterize the parameter dependence of the localization length, as this is crucial for the electrical properties in coherent transport
through the system.

In shorter wires ($L<\ell_{\it loc}$) transport is diffusive and typical of metallic conductors characterized by Ohm's law. In this case there is a linear relation 
between the electrical resistance and total length  $L$. That is, in our model, we may expect a regime such that
\begin{equation}
\label{eq13b}
\left\langle\frac{1}{T}\right\rangle
= 
\frac{1}{N_p}+\frac{1}{N_p\ell_\Omega}\langle L\rangle\; ,
\end{equation}
where the averages are regarding system realizations and we have defined a diffusive (ohmic) length $\ell_\Omega$ that characterizes the electron mean free 
path. The constant contribution $1/N_p$ ($N_p$ number of propagating modes) 
in Eq.\ (\ref{eq13b}) represents the contact resistance, present even without any scattering effect.
Localization length and mean free path are actually related by  $\ell_{\it loc}\approx N_p\ell_\Omega$,
relation that we have explicitly checked for segmented wires (see also Ref.\ \cite{Garcia97}).
For completeness, besides the localized and diffusive regimes the so-called ballistic regime corresponds to $L<<\ell_\Omega$, such that 
only the contact resistance contribution matters in Eq.\ (\ref{eq13b}).

Yet another characteristic length may be obtained using a semiclassical approximation.
In a semiclassical description the scatterers representing the vertices add up their effects incoherently, yielding a total transmission $T_{sc}$ that 
is independent of the segments length $\ell_0$,
\begin{equation}
\label{eq14}
\frac{1}{T_{sc}}-\frac{1}{N_p}
=
\sum_{i=1}^{N_v}\left(
\frac{1}{T^{(i)}}-\frac{1}{N_p}
\right)\;,
\end{equation}
where $T^{(i)}\equiv|t^{(i)}|^2$ is the transmission probability corresponding to vertex $i$.
As the mean total length is $\langle L\rangle \approx N_v \ell_0$, Eq.\ (\ref{eq14}) yields an
ohmic scaling similar to Eq.\ (\ref{eq13b}), 
\begin{equation}
\frac{1}{T_{sc}} = \frac{1}{N_p} + \frac{1}{\ell_{sc}}\langle L\rangle \;,  
\end{equation}
where we defined a {\it semiclassical length} 
\begin{equation}
\label{eqlsc}
\ell_{sc}= \frac{\ell_0}{
\left(\frac{1}{N_v}\sum_i{\frac{1}{T^{(i)}}}\right)
-
\frac{1}{N_p}
}\; .
\end{equation}
As shown below, $\ell_{sc}$ yields an estimate of $\ell_{\it loc}$ that averages all possible oscillations and resonances due to quantum interference.

\begin{figure}[t]
\centering
\resizebox{0.4\textwidth}{!}{
	\includegraphics{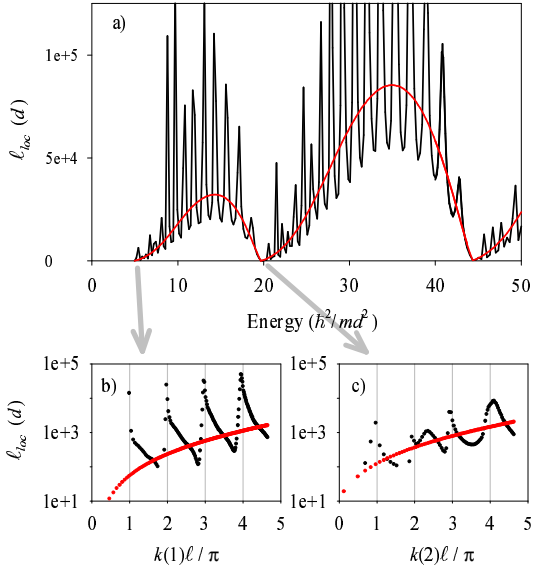}
}
\caption{Localization length as a function of energy. The lower panels 
show the dependence on mode wave number for the energy ranges
around the first and second onsets for mode 
activation.
The red line is the semiclassical result $\ell_{sc}$ defined in Eq.\ (\ref{eqlsc}).
Parameters: $\ell_0=10 d$, $R_0=0.2 d$, $p_\ell=0.01$, $p_R=0.01$, $\alpha_0=60^\circ$.
}
\label{F2}
\end{figure}

\begin{figure}[t]
\centering
\resizebox{0.4\textwidth}{!}{
	\includegraphics{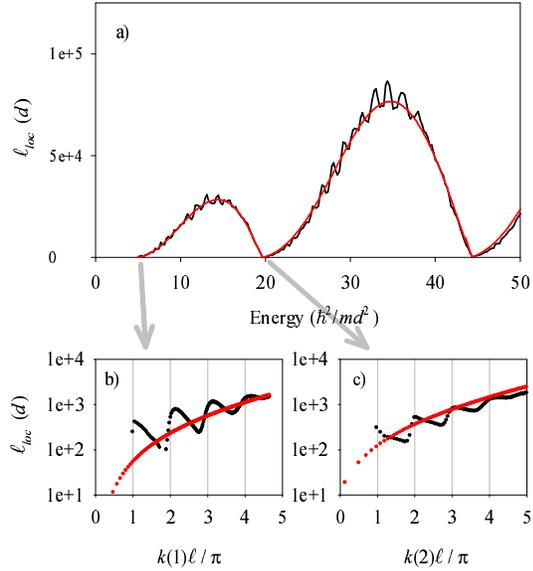}
}
\caption{Same as Fig.\ \ref{F2} but with $p_\ell=0.1$.
}
\label{F3}
\end{figure}

The ohmic regime is sometimes referred to as diffusive or 
metallic
and it is characterized by a relatively low transmission, as compared to the maximum value allowed by the conductance quantization of the channel.
Within the random matrix theory  this is a regime of
universal conductance fluctuations \cite{Beenakker97},
the transmission being normally distributed with a statistical 
dispersion $\Delta T=\sqrt{2/15}$ (in systems with time reversal symmetry like ours).
On the other hand, in the ballistic limit the transmission reaches the quantized maximum values allowed by the number of propagating modes $N_p$.
We expect this regime only in short-enough wires and relatively high energies, such that scattering
effects become negligible.

\begin{figure}[t]
\centering
\resizebox{0.4\textwidth}{!}{
	\includegraphics{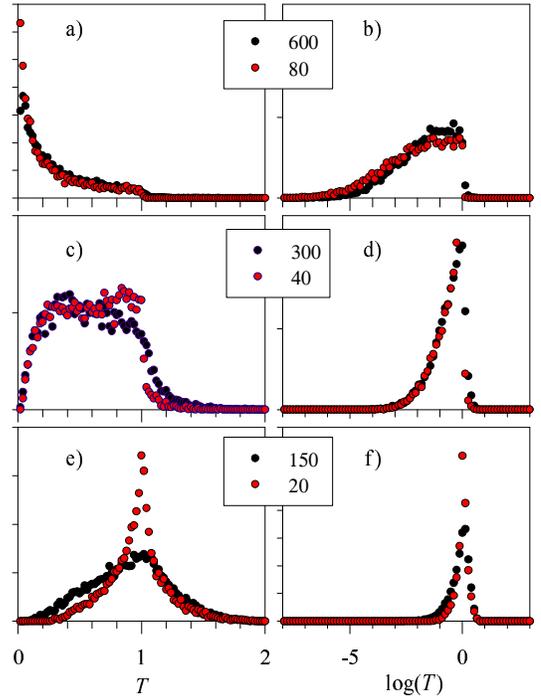}
}
\caption{Probability distribution of $T$ (left) and $\log(T)$ (right) in the crossover regime.
The number of segments corresponding to each symbol varies in each panel
as indicated.
Black circles correspond to an energy $E=19.8 \hbar^2/md^2$ for which the localization length
is $\ell_{\it loc}
\approx 3000d$. On the other hand, red circles correspond to $E=20.1 \hbar^2/md^2$
and $\ell_{\it loc}\approx280d$. The rest of parameters are as in Fig.\ \ref{F2}.
}
\label{Fn}
\end{figure}

\section{Results}

Figure \ref{F2} shows the energy dependence of $\ell_{\it loc}$ in a segmented wire.
A conspicuous resonant behavior is seen, with closely lying spikes and an overall  beating pattern 
corresponding to the successive activation of propagating modes.
The beating is accurately reproduced by the semiclassical length $\ell_{sc}$ (red line)
that nicely averages the resonant oscillations. The resonances occur when
an integer number of wave lengths fit in the segment length $\ell_0$, i.e.,
$k(i)\ell_0 = n\pi$, with $n$ an integer.
This resonant condition does not depend on the vertex parameters $\alpha$, $R_0$, neither on $p_R$, but it quickly degrades when $p_\ell$ increases,
as seen in Fig.\ \ref{F3}. 
As shown in this figure, a dispersion of $\pm 10\%$ is 
enough to greatly reduce the resonance peaks. 
The resonant behavior with one propagating mode is 
qualitatively similar to the behavior in strictly 1D systems
discussed in Refs.\ \cite{Herrera13,Diaz15}. 
We stress, however, that we also find similar resonances in 
higher energy regions, where more modes become propagating.

The crossover between diffusive and localized regimes of disordered wires is known to be 
characterized by a nontrivial evolution of the $T$ and $\log(T)$ distributions 
\cite{Muttalib99,Garcia01,Gopar02,Alcazar13}. 
We have explored whether  the crossover is greatly affected by the resonant condition or not.
More specifically, we choose parameter sets corresponding to a maximum and a minimum in 
localization length and check the evolution with varying number of segments. Figure \ref{Fn}
shows that even when the localization length changes by more than an order of magnitude
with a small energy change (spiking behavior in Fig.\ \ref{F2}), the qualitative evolution 
of the probability distributions is very similar. 

The results of Fig.\ \ref{Fn}a,b correspond to 
$L\approx2\ell_{\it loc}$, just entering the localized regime. They show  a long-tail $T$ distribution with a change of behavior at $T=1$. On the other hand, $\log(T)$ is given by an asymmetric Gaussian
in this region. The central panels, Figs.\ \ref{Fn}c,d correspond to the middle of the crossover 
with $L\approx \ell_{\it loc}$ and show a rather flat distribution 
of transmissions with a dip at $T\approx 0$ \cite{Wang98,Plerou98}. The lower panels Figs.\ \ref{Fn}e,f signal the beginning of the diffusive regime $L\approx 0.5 \ell_{\it loc}$ and show 
a kink at $T=1$ separating two regions in the $T$-distribution. 
These crossover features are already known and they agree well with the results of studies of 
disordered wires \cite{Muttalib99,Garcia01,Gopar02}.
They are, nevertheless, shown here to stress the similar evolution
for widely different localization lengths due to resonance.
It is worth mentioning that when the localization length is 
very short, as for the red dots in Fig.\ \ref{Fn}, the transition from 
quasi-ballistic to localized is more abrupt, leaving a quite reduced diffusive range. We attribute to this enhanced quasi-ballistic behavior 
the increased kink at $T\approx 1$ of the red dots in Fig.\ \ref{Fn}e, as compared with the black ones.

\begin{figure}[t]
\centering
\resizebox{0.4\textwidth}{!}{
	\includegraphics{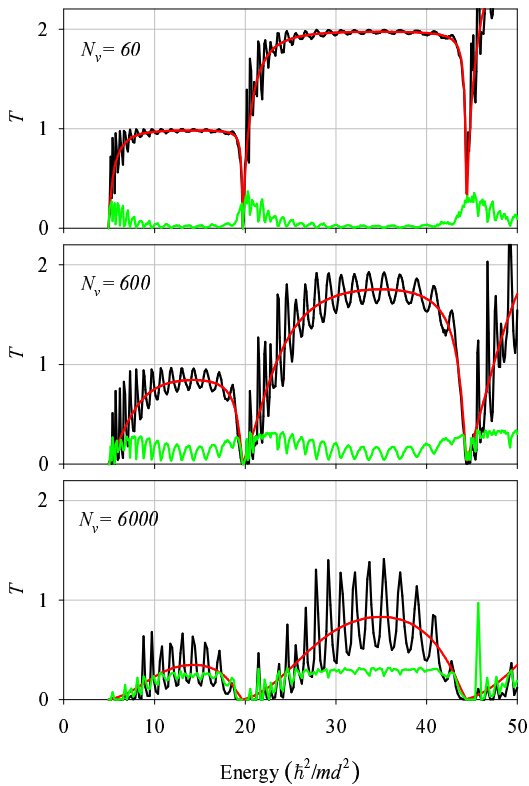}
}
\caption{Mean conductance as a function of energy for increasing number
of segments as indicated in each panel. The same parameters of Fig.\ \ref{F2} have been used.
The green line shows the dispersion of the statistical distribution of transmissions while the red line
is the semiclassical result $T_{sc}$. 
}
\label{F4}
\end{figure}

\begin{figure}[t]
\centering
\resizebox{0.4\textwidth}{!}{
	\includegraphics{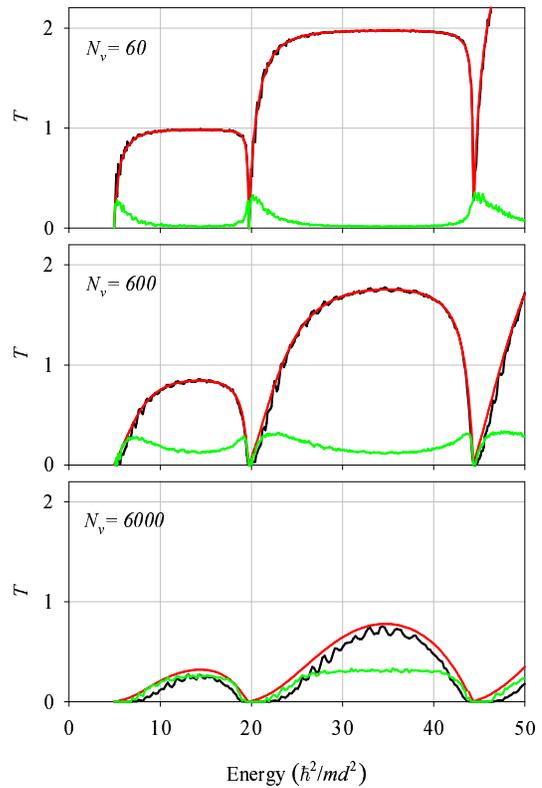}
}
\caption{
Same as Fig.\ \ref{F4} but with $p_\ell=0.1$.}
\label{F5}
\end{figure}

We discuss next the mean conductance of a wire with a fixed number 
of segments (Figs.\ \ref{F4} and \ref{F5}). The conductance shows a
general tendency to increase in discrete steps as the energy increases,
typical of quantum wires. The conductance plateaus, however, are 
distorted in remarkable ways.
First we notice in Fig.\ \ref{F4} the already mentioned resonant oscillations, present at the beginning of each plateau for the smaller
$N_v$ and eventually extending to all the plateau for the larger $N_v$'s.
A pronounced conductance dip is also observed at the beginning of each plateau. For the shorter wires the plateaus saturate at the quantized values,
while in the longer ones there is no clear saturation and the transmission is
in general much lower than the corresponding quantized values.

The physics implied by Fig.\ \ref{F4} can be understood as a typical evolution of a finite-wire conductance
with increasing energy: from a localized regime near 
the plateau onset, to an ohmic (diffusive) regime and eventually 
reaching a ballistic 
regime if the wire is short enough. 
The localized regime occurs at the plateau onset, where
the localization length is small (Figs.\ \ref{F2} and \ref{F3}).
As the energy increases, an ohmic regime is reached, characterized 
by a sizeable dispersion of the transmissions and by the linearity 
of the inverse transmission with length.
In short wires (Fig.\ \ref{F4} upper panel) the system may also reach 
the quantum ballistic regime,
with a quantized unitary transmission and vanishing dispersion.

As with the localization lengths, 
the strong transmission oscillations due to wave number quantization 
are quenched if the segment lengths vary by a sizeable amount. Figure \ref{F5} shows the transmission with $p_\ell=0.1$. In this case, there is 
a better correspondence with the semiclassical result.

\section{Conclusions}

A model of random segmented 2D wire with circular
bends has been presented. We focussed on the scattering induced by the bends and how this leads to the emergence of localization.
A strong resonant behavior is predicted when the segments are all of very similar lengths. A spiking behavior of the localization length is found,
not only with a single propagating mode, but also in presence of several
modes. A beating pattern of the spiking is accurately reproduced by a semiclassical model, averaging quantum oscillations and resonances.
The localization resonances are reduced when the distribution of segment lengths 
gets broader.

The localized-diffusive crossover is shown to agree with the known 
behaviors from disordered wires. The same qualitative evolution  
of the $T$ and $\log(T)$ distributions is found for large and small 
localization lengths, moving across the resonance spikes. For short localization lengths the diffusive regime is much reduced, yielding a more abrupt
evolution from ballistic to localized cases. A fixed-length wire 
typically
evolves with increasing energy from localization at the beginning 
of each transmission plateau, to a diffusive regime and to 
ballistic behavior towards the plateau end. The ballistic regime may be reached only if the wire is short enough.

\begin{acknowledgments}
This work was funded by MINECO-Spain (grant FIS2014-52564),
CAIB-Spain (Conselleria d'Educaci\'o, Cultura i Universitats) and 
FEDER. 
\end{acknowledgments}

\bibliographystyle{apsrev4-1}
\bibliography{andrefs}

\end{document}